\documentclass[aps,prl,twocolumn]{revtex4}

\usepackage{graphicx,latexsym,color}

\def\be{\begin{equation}}

\def\ee{\end{equation}}

\def\bea{\begin{eqnarray}}

\def\eea{\end{eqnarray}}

\def\bi{\begin{itemize}}

\def\ei{\end{itemize}}

\begin{document}
%
\title{Gauss sum factorization with cold atoms}
\author{M. Gilowski$^{1}$, T. Wendrich$^{1}$, T. M{\"u}ller$^{1}$, Ch. Jentsch$^{2}$, W.~Ertmer$^{1}$, E.~M.~Rasel$^{1}$, and W.~P.~Schleich$^{3}$}                     
\affiliation{$^{1}$Institut f\"{u}r Quantenoptik, Leibniz
Universit\"{a}t
Hannover, Welfengarten 1, D-30167 Hannover, Germany\\
$^{2}$Astrium GmbH - Satellites, 88039 Friedrichshafen, Germany\\
$^{3}$Institut f{\"u}r Quantenphysik, Universit{\"a}t Ulm,
Albert-Einstein-Allee 11, D-89081 Ulm, Germany}

\date{Received: date / Revised version: date}
\begin{abstract}
We report the first implementation of a Gauss sum factorization
algorithm by an internal state Ramsey interferometer using cold
atoms. A sequence of appropriately designed light pulses interacts
with an ensemble of cold rubidium atoms. The final population in the
involved atomic levels determines a Gauss sum. With this technique we
factor the number $N$=263193.
\end{abstract}
\maketitle
\par
The Shor algorithm~\cite{Shor} to factor numbers and its
NMR-implementation~\cite{NMRimplementation} have propelled the
field of quantum computation~\cite{QC}. Recently a different
factorization scheme~\cite{Merkel} taking advantage of the
periodicity properties of Gauss sums~\cite{Maier} has been
proposed~\cite{Clauser} and verified by two
NMR-experiments~\cite{Mehring,Mahesh} and one experiment based on
short laser pulses~\cite{Girard}. In the present paper we report
the first implementation of Gauss sum factorization based on
matter-wave interferometry~\cite{Kasevich} with cold rubidium
atoms and use it to find the factors of $N=263193$.

Our method rests on the observation that the truncated Gauss sum
\begin{equation}\label{eqn:total_Signal}
\mathcal{C}^{(M)}_{N}(l)=\frac{1}{M+1}\sum_{m=0}^{M}cos\left(2\pi
m^{2}\frac{N}{l}\right)
\end{equation}
consisting of $M+1$ terms yields unity if the integer $l$ is a factor
of $N$. In contrast, for integer non-factors $l$ destructive
interference leads to a small value of $\mathcal{C}^{(M)}_{N}(l)$.
Thus the algorithm checks if a given integer $l$ is a factor of $N$
or not. This testing is performed by a quantum system and its time
evolution determines if a trial factor is a factor or not. Since in
the worst case we have to test all prime numbers up to $\sqrt{N}$ the
prime number theorem~\cite{Maier} predicts the upper bound
$\sqrt{N}/$log$N$ for the number of trials. As a consequence our
method in the present form scales exponentially which is not
surprising since it does not involve entanglement yet and relies
solely on interference. In this respect, our technique is very much
in line with the recent critical discussion~\cite{Mermin} of the
connection between quantum mechanics and factorizing.

Our procedure for implementing Gauss sums is reminiscent of the
method used in Ref.~\cite{Mehring}. Common to both techniques is a
sequence of pulses which imprints on a two-level quantum system a
sequence of well-defined phases. For appropriately chosen pulses the
excitation probability takes the form of a Gauss sum.
\begin{figure}[t]
\resizebox{\columnwidth}{!} {
  \includegraphics{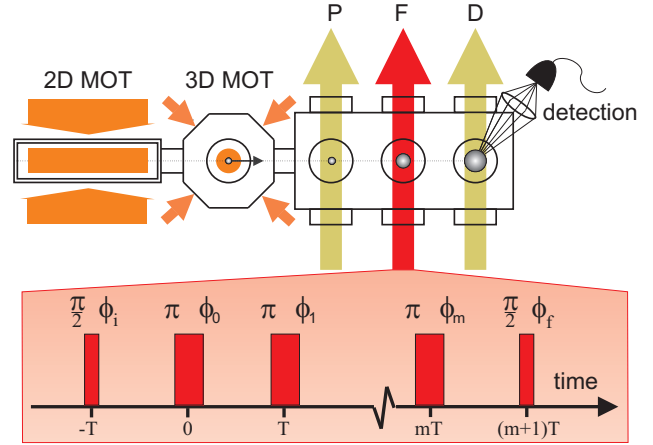}
}\caption{Implementation of the Gauss sum factorization algorithm
using cold rubidium atoms. They are launched by a double MOT system
and prepared in the region \emph{P} by an appropriate pulse sequence
in the atomic ground state. In the factorization zone \emph{F} the
atoms interact with a sequence of pulses driving a hyperfine
transition. We start with a $\pi/2$-pulse followed by a sequence of
$m+1$ $\pi$-pulses with the phase $\phi_{k}(l)$ and conclude by
another $\pi/2$-pulse. The phase during the $\pi/2$-pulses is
$\phi_{i}=\phi_{f}=-90^{\circ}$. A fluorescence detection in the
region \emph{D} measures the populations in both states which
determines the interference signal $c_{m}(l)$. We record $c_{m}$ for
$0\leq m\leq M$ and the sum over $c_{m}$ yields the Gauss sum
$\mathcal{C}^{(M)}_{N}(l)$.} \label{fig:setup}
\end{figure}

Despite these similarities our technique is different in many
aspects: (\emph{i}) In contrast to the
NMR-approaches~\cite{Mehring,Mahesh} we do not use liquids but an
ensemble of cold atoms. The spectacular control of internal as well
as external atomic degrees of freedom with the help of lasers, opens
up a new avenue towards factorization. Indeed, rubidium atoms feature
two long-living hyperfine ground states which can be coherently
manipulated by a two-photon Raman-transition~\cite{Kasevich2}.
Moreover, the use of cold atoms provides us with long interaction
times and the possibility of a large number of pulses. (\emph{ii}) In
our approach there are no projection measurements between the
pulses~\cite{Mehring}. The measurement takes place after completing
the sequence of pulses. (\emph{iii}) The Gauss sum results from the
addition of the measured signals obtained from an increasing number
of pulses. As a consequence our resources in pulses scale
quadratically whereas in the NMR-approach they only scale linearly.
Despite this unfavorable scaling of the present technique we maintain
that cold atoms offer several advantages. Most importantly, they have
already proven to be ideal objects to be entangled. In particular,
two-qubit quantum gates relying on cold atoms in standing light
waves~\cite{Bloch} have been realized. Therefore, we consider our
experiment as a stepping stone towards more complex arrangements
including entangled quantum systems.

Our experimental setup shown in Fig.~\ref{fig:setup} is part of an
atom interferometer~\cite{Jentsch} designed to measure rotations
with a high precision. The source of the cold Rb-atoms is a double
stage MOT~\cite{Raab,Mueller}. A two-dimensional trap creates an
atomic beam with a flux of $5\times10^{9}$~at/s for the efficient
loading of a subsequent 3D-MOT. Approximately $N_{at}=10^{8}$
trapped atoms are launched in a moving molasses, similar to atomic
fountains~\cite{Bize} and are transferred into the atomic ground
state $|5^{2}S_{1/2},F=1,m_{F}=0>$ using a multi-stage
preparation.

In order to implement the Gauss sum, we use the pulse sequence
illustrated in Fig.~\ref{fig:setup}. An initial $\pi/2$-pulse with
phase $\phi_{i}=-90^{\circ}$ prepares a coherent superposition of
ground and excited state. The latter corresponds to the hyperfine
ground state $|5^{2}S_{1/2}$, $F=2,m_{F}=0>$. This superposition is
most sensitive to the phases of $\pi$-pulses. Indeed, after a time
$T$ we apply the factorization sequence consisting of $m+1$
$\pi$-pulses each separated by the time $T$. The $k$-th pulse has the
phase~\cite{Mehring}
\begin{eqnarray}
\phi_{k}(l)\equiv\left\{
\begin{array}{ccc}
0 & \rm{for} & k=0\\a_{k}(N)/l & \rm{for} & 1\leq k\\
\end{array}\right.
\end{eqnarray}
with $a_{k}(N)\equiv(-1)^{k}\pi N(2k-1)$ which induces a coherent
evolution of the atomic ensemble. We conclude at the time $(m+1)T$
the factorization sequence by a $\pi/2$-pulse with phase
$\phi_{f}=-90^{\circ}$ to convert the phase evolution into a
population difference between the two atomic states which is measured
by a state-selective fluorescence detection~\cite{Sortais}.

The populations in the two states are governed~\cite{Merkel,Mehring}
by the interference signal
\begin{equation}\label{eqn:S_total}
c_{m}(l)\equiv cos\left(2\pi m^{2}\frac{N}{l}\right)
\end{equation}
which assumes values between -1 and +1 corresponding to all atoms in
the ground or excited state, respectively. When we repeat the pulse
sequence for increasing $m$ with $0\leq m\leq M$ and add the
interference signals we arrive after normalization at the total
signal $\mathcal{C}^{(M)}_{N}(l)$ determined by
Eq.~(\ref{eqn:total_Signal}).

The multi-pulse excitation is executed by two digitally phase-locked
Raman-lasers~\cite{oPLL} which drive the hyperfine transition at
approximately 6.834~GHz. The two beams are co-propagating through the
factorization zone perpendicular to the trajectories of the atoms. In
contrast to the velocity-sensitive excitation in inertial atomic
sensors~\cite{Canuel}, in the present velocity-insensitive
configuration we can neglect phase contributions from inertial
forces.

A low phase noise oscillator serves as the reference for the phase
locking setup. The phase $\phi_{k}(l)$ determined in advance by a
computer for each trial factor $l$ is adjusted electronically by a
synthesizer. The stringent requirements on the phase control in atom
interferometry make the realization of the factorization experiment
possible. The total phase error of the Raman-laser system is
approximately 1~mrad. The length of a $\pi$-pulse is approximately
23~$\mu$s while the time $T$ between two $\pi$-pulses is 100~$\mu$s
which is sufficient for the electronic control loop to adjust the
required phase in the laser system.

\begin{figure}[b]
\resizebox{\columnwidth}{!} {
  \includegraphics{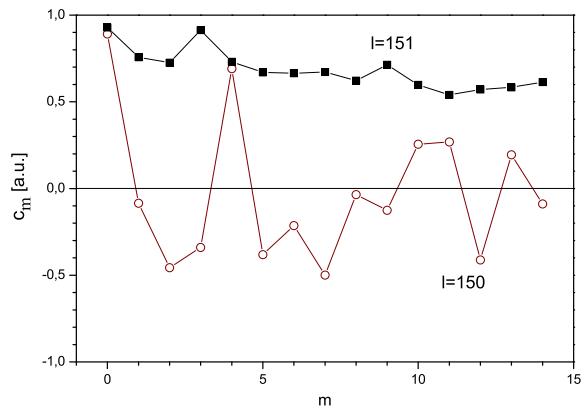}
} \caption{Measured interference signals $c_{m}(l)$ as a function of
the summation index $m$ of the Gauss sum Eq.~(\ref{eqn:total_Signal})
determined by the number $m+1$ of pulses in the factorization
sequence for $N=263193=3\times7\times83\times151$. Whereas $c_{m}(l)$
is approximately constant for the factor $l$=151 leading to the value
$\mathcal{C}^{(14)}_{N}(151)$=0.69 the corresponding signal for the
non-factor $l$=150 oscillates creating the small value
$\mathcal{C}^{(14)}_{N}(150)$= --0.02. Here we have used the maximum
number $M+1=15$ of pulses.} \label{fig:zwei_l}
\end{figure}
We now turn to the discussion of the results of our factorization
experiment exemplified by the number
$N=263193=3\times7\times83\times151$. Figure~\ref{fig:zwei_l} brings
out most clearly that the dependence of the interference signal
$c_{m}(l)$ on $m$ is dramatically different for a factor of $N$ such
as $l=151$ and a non-factor such as $l=150$. Since for a factor
$c_{m}(l)$ is approximately constant as a function of $m$ the
constructive addition of the terms in the definition
Eq.~(\ref{eqn:total_Signal}) of the Gauss sum leads to a total signal
$\mathcal{C}_{N}^{(M)}(l)$ close to unity. On the other hand a
non-factor produces an oscillatory function $c_{m}(l)$ which takes on
values between -1 and +1. Thus the total signal
$\mathcal{C}_{N}^{(M)}(l)$ for a non-factor is rather small.

\begin{figure}[t]
\resizebox{\columnwidth}{!} {
  \includegraphics{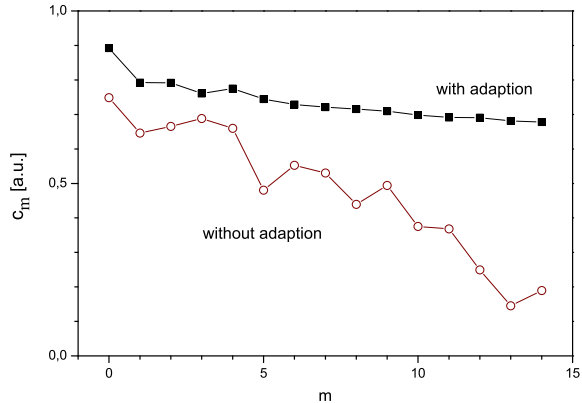}
} \caption{Average of the interference signals $c_{m}(l)$
corresponding to three factors of $N=263193$ as a function of $m$
with and without a parabolic adaption of the pulse length.}
\label{fig:parabola}
\end{figure}
Figure~\ref{fig:zwei_l} also shows a decay of the interference
signal with increasing $m$. This behavior originates from the
fact, that the atomic cloud on its trajectory sees a Gaussian
intensity distribution of the Raman-laser beams which translates
into a Gaussian distribution of the lengths of the $\pi/2$- and
$\pi$-pulses. This leads to a reduction of the transition
probability and to a decrease of the interference signal during
the multi-pulse sequence.

We have compensated this disturbance by appropriately adjusting
the pulse length over the total interaction region. For this
purpose we have measured the lengths of the $\pi$-pulses for atoms
in the center and in the edge of the interaction region which is
20~$\mu s$ and 26~$\mu s$, respectively. Since we can approximate
a Gaussian in the neighborhood of its center by a polynomial of
second degree we can connect the two measured pulse lengths by a
parabola which provides us with an approximation of the pulse
lengths over the whole interferometer region.

In Fig.~\ref{fig:parabola} we compare the interference signals with
and without such a pulse length adaption in their dependence on $m$.
For this purpose we concentrate on factors only. In order to obtain a
smooth curve we take the average of three such curves each
corresponding to a different factor. This procedure clearly shows
that with the adaption the decay is slowed down and consequently we
can apply more pulses.

The remaining decay of $c_{m}(l)$ results from the interaction of
the atomic cloud which has currently a diameter of about 5~mm in
the interaction region with the Gaussian intensity distributed
Raman-lasers leading to a reduction of the transition probability
during each single pulse. Thus lower temperatures of the atoms
would slow down the decay of $c_{m}(l)$.

An upper limit for the number $M+1$ of possible pulses in the
factorization sequence is experimentally set by the total interaction
time $(M+2)T$ which results from the non-vanishing velocity
$v_{at}\sim~$4.4~m/s of the atoms propagating through the interaction
region given by the diameter $d\sim~$30~mm of the Raman-laser beams.
Thus we can apply up to 20 pulses without any excessively disturbing
effects.

\begin{figure}[t]
\resizebox{\columnwidth}{!} {
  \includegraphics{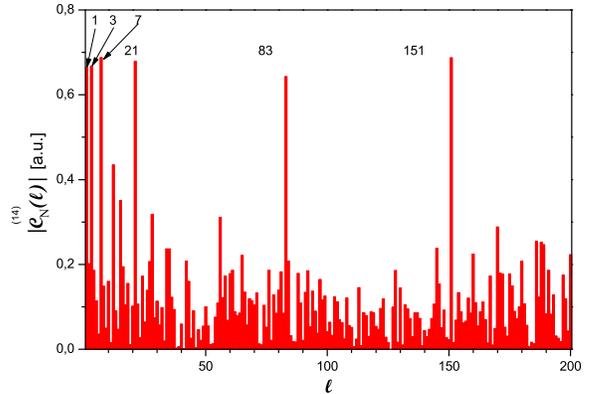}
} \caption{Experimental Gauss sum factorization based on cold atoms
and exemplified by the number $N=263193=3\times7\times83\times151$
using the maximum number $M+1$=15 of pulses. Here we display the
absolute value of the sum $\mathcal{C}^{(14)}_{N}(l)$ over the
interference signals $c_{m}(l)$, that is the measured Gauss sum as a
function of the trial factors $1\leq l\leq200$. Dominant peaks
correspond to factors of $N$. The background is produced by the
non-factors.} \label{fig:fak}
\end{figure}
\begin{figure}[b]
\resizebox{\columnwidth}{!} {
  \includegraphics{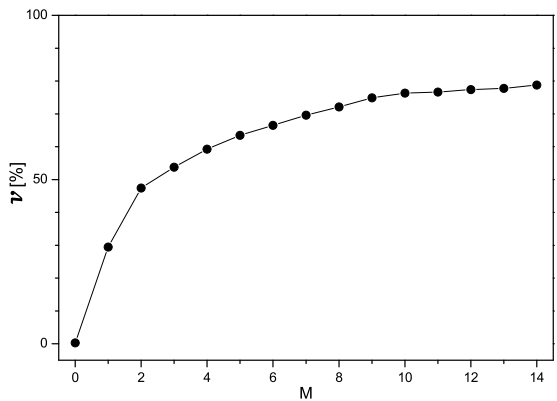}
}\caption{Measured contrast $\mathcal{V}$ of the factorization
pattern for $N$=263193 as a function of $M$. Here $M+1$ is the
maximum number of factorization pulses. Already six pulses ($M=5$)
yield a contrast larger than 60\%.} \label{fig:kontrast}
\end{figure}
Figure~\ref{fig:fak} demonstrates the successful implementation of a
Gauss sum factorization algorithm by an internal state Ramsey
interferometer using cold atoms. Here we display the absolute value
of the sum $\mathcal{C}^{(M)}_{N}(l)$ of the interference signals
$c_{m}(l)$ for all analyzed integer trial factors $l$ for the number
$N=263193=3\times7\times83\times151$ recorded with the maximum number
$M+1=15$ of pulses. The corresponding factors of $N$ stand out
clearly from the background created by the non-factors. We also
obtain factor-like signals for products of factors as exemplified by
$l$=21, in complete agreement with the Gauss sum,
Eq.~(\ref{eqn:total_Signal}).

Figure~\ref{fig:kontrast} displays the experimentally obtained
contrast $\mathcal{V}$ of a factorization pattern such as the one in
Fig.~\ref{fig:fak} as a function of the number $M$. Following
Ref.~\cite{Mehring} we have defined $\mathcal{V}$ as the ratio
between the difference and the sum of the measured averaged absolute
values of the Gauss sum $\mathcal{C}^{(M)}_{N}(l)$ at factors and
non-factors. We find that with just six pulses corresponding to $M=5$
the number $N=263193$ can be factored with a contrast larger than
60\%, in agreement with the theoretical prediction of
Ref.~\cite{Mehring}.

We have demonstrated the first implementation of a Gauss sum
algorithm to factor numbers based on cold atoms and have
successfully factorized the number $N$=263193. Our results are
comparable to those of the NMR-experiments~\cite{Mehring,Mahesh}.
However, the use of cold atoms not only represents an alternative
approach but also allows us to envision several extensions:
(\emph{i}) Quadratic phases can be generated without prior
calculation by a computer by a linear sweep of an external
magnetic or electric field during the multi-pulse
sequence~\cite{Merkel}. (\emph{ii}) The preparation of the atomic
ensemble in an optical lattice opens up the possibility of
applying an almost arbitrarily large number of pulses.
(\emph{iii}) A Gauss sum factorization scheme involving
entanglement could rely on a large number of entangled atoms each
one located in the minima of an optical lattice~\cite{Bloch2}
providing us with a massive parallelism. The experiment reported
in the present paper is the first step in these directions.

We appreciate stimulating discussions with B.~Girard, D.~Haase and
M.~$\breve{\textnormal{S}}$tefa$\breve{\textnormal{n}}\acute{\textnormal{a}}$k.
The work is supported by the SFB 407 of the Deutsche
Forschungsgemeinschaft and the FINAQS cooperation of the European
Union. One of us (WPS) is grateful to the Max-Planck Society and the
Alexander von Humboldt Stiftung for their support. Moreover, he
acknowledges the support by the Ministerium f{\"u}r Wissenschaft und
Kunst, Baden-W{\"u}rttemberg and the Landesstiftung
Baden-W{\"u}rttemberg in the framework of the Quantum Information
Highway A8 and the Center for Quantum Engineering.


\begin{thebibliography}{}
\bibitem{Girard} B.~Girard has kindly informed us about his recent experiment with B.~Chatel and D.~Bigourd factorizing numbers with Gauss sums based on a sequence of femtosecond laser pulses generated by a pulse shaper. The two-level atom is replaced by a spectrometer tuned to a single wavelength. The pulses are delayed such that the relative phase of a given wavelength is the required phase for the individual terms of the Gauss sum.
\end{thebibliography}
\end{document}